\documentclass[english,12pt]{article}
\usepackage[T1]{fontenc}
\usepackage[latin1]{inputenc}
\usepackage{geometry}
\usepackage{epsfig}
\usepackage{xspace}
\usepackage{axodraw}
\usepackage{cite}

\IfFileExists{url.sty}{\usepackage{url}}
                      {\newcommand{\url}{\texttt}}

\makeatletter

\usepackage{babel}
\makeatother
\date{}
\title{\huge \bf On the Rational Terms of the one-loop amplitudes
\\[1cm]}

\author{
{\bf
 Giovanni Ossola$^{(1)}$ ,
 Costas G.~Papadopoulos$^{(1)}$ ,
 Roberto Pittau$^{(2,3)}$
}\\\\
{\normalsize $^{(1)}$Institute of Nuclear Physics, NCSR Demokritos,
15310 Athens, Greece}\\\\
{\normalsize $^{(2)}$Departamento de F\'{i}sica Te\'orica y del Cosmos}\\
{\normalsize Centro Andaluz de F\'{i}sica de Part\'{i}culas Elementales (CAFPE)}\\
{\normalsize Universidad de Granada, E-18071 Granada, Spain} \\\\
{\normalsize $^{(3)}$Dipartimento di Fisica Teorica}\\
{\normalsize Univ. di Torino and INFN sez. di Torino}\\
{\normalsize V. P. Giuria 1, I-10125 Torino, Italy}\\[1cm]
}
\begin{document}
\newcounter{im}
\setcounter{im}{0}
\newcommand{\exampleSp}{\stepcounter{im}\includegraphics[scale=0.9]{SpinorExamples_\arabic{im}.eps}}
\newcommand{\myindex}[1]{\label{com:#1}\index{{\tt #1} & pageref{com:#1}}}
\renewcommand{\topfraction}{1.0}
\renewcommand{\bottomfraction}{1.0}
\renewcommand{\textfraction}{0.0}
\newcommand{\nc}{\newcommand}
\nc{\eqn}[1]{Eq.~\ref{eq:#1}}
\nc{\be}{\begin{equation}}
\nc{\ee}{\end{equation}}
\nc{\ba}{\begin{array}}
\nc{\ea}{\end{array}}
\nc{\bea}{\begin{eqnarray}}
\nc{\eea}{\end{eqnarray}}
% begin my commands
\nc{\bqa}{\begin{eqnarray}}
\nc{\eqa}{\end{eqnarray}}
\newcommand{\nl}{\nonumber \\}
\def\db#1{\bar D_{#1}}
\def\zb#1{\bar Z_{#1}}
\def\d#1{D_{#1}}
\def\tld#1{\tilde {#1}}
\def\slh#1{\rlap / {#1}}
\def\eqn#1{Eq.~(\ref{#1})}
\def\eqns#1#2{Eqs.~(\ref{#1}) and~(\ref{#2})}
\def\eqnss#1#2{Eqs.~(\ref{#1})-(\ref{#2})}
\def\fig#1{Fig.~{\ref{#1}}}
\def\figs#1#2{Figs.~\ref{#1} and~\ref{#2}}
\def\sec#1{Section~{\ref{#1}}}
\def\app#1{Appendix~\ref{#1}}
\def\tab#1{Table~\ref{#1}}

\newcommand{\bfig}{\begin{center}\begin{picture}}
\newcommand{\efig}[1]{\end{picture}\\{\small #1}\end{center}}
\newcommand{\flin}[2]{\ArrowLine(#1)(#2)}
\newcommand{\wlin}[2]{\DashLine(#1)(#2){2.5}}
\newcommand{\zlin}[2]{\DashLine(#1)(#2){5}}
\newcommand{\glin}[3]{\Photon(#1)(#2){2}{#3}}
\newcommand{\lin}[2]{\Line(#1)(#2)}
\newcommand{\sof}{\SetOffset}

% end my commands

\maketitle
\begin{abstract}
The various sources of Rational Terms contributing to the one-loop
amplitudes are critically discussed. We show that the terms
originating from the generic $(n-4)-$dimensional structure of the
numerator of the one-loop amplitude can be derived by using
appropriate Feynman rules within a tree-like computation. For the
terms that originate from the reduction of the $4-$dimensional part
of the numerator, we present two different strategies and explicit
algorithms to compute them.

%, together with a few strategies and explicit algorithms to compute
%them.
% , within and outside the
%framework of the {\tt OPP} method.
\end{abstract}

\newpage
\vspace*{\fill}
\tableofcontents
\vspace*{\fill}
\newpage

% main text
\section{Introduction}
In the last few years, a big effort has been devoted by several authors
to the problem of computing one-loop amplitudes efficiently~\cite{Review}.
Besides Standard techniques, where tensor reduction/computation is performed,
numerically or analytically, new developments emerged, originally
inspired by unitarity arguments (the so called unitary and generalized unitarity methods)
%~\cite{unitarity-cut}~\cite{Britto:2004nc}~\cite{twistors}
%~\cite{Anastasiou},
~\cite{unitarity-cut} -\cite{Anastasiou},
 in which the tensor reduction/computation
is substituted by the problem of determining the {\em coefficients}
of the contributing scalar
one-loop functions. This possibility relies on the fact that the
basis of one-loop integrals is known in terms of Boxes, Triangles,
Bubbles and (in massive theories) Tadpoles, in such a way
that, schematically, one can write
a Master Equation for any one-loop amplitude ${\cal M}$ such as:
\bqa
{\cal M}=
\sum_i d_i {\rm ~Box}_i
+\sum_i c_i {\rm ~Triangle}_i
+\sum_i b_i {\rm ~Bubble}_i
+\sum_i a_i {\rm ~Tadpole}_i
+ {\rm R}\,,
\eqa
where $d_i$, $c_i$, $b_i$ and $a_i$ are the coefficients
to be determined.

Nevertheless, in practice, only the part of the amplitude
proportional to the one-loop scalar functions can be obtained
straightforwardly in the unitary cut method.
The remaining Rational Terms, RTs, (${\rm R}$ in the above equation) should
be reconstructed by other means, either by direct computation~\cite{directcomp1,directcomp2}
or by boostrapping methods~\cite{boostrap}.
In~\cite{Giele:2008ve} the RTs are obtained by explicitly computing the
amplitude at different integer value of the space-time dimensions.
On the other hand, in another recently proposed method, {\tt OPP} \cite{Ossola:2006us}, a class
of terms contributing to ${\rm R}$ can be naturally derived in the same
framework used to determine all the other coefficients.

In this paper, we critically analyze the various
sources of RTs appearing in one-loop amplitudes, by classifying them in
two categories: ${\rm R= R_1+R_2}$, also presenting
a few computational methods.
In the next section, we investigate the origin of ${\rm R_2}$ and
develop a practical computational strategy.
 In Section \ref{strategy1}, after a brief recall of the {\tt OPP} method, we
present a way to compute ${\rm R_1}$, strictly connected to the
{\tt OPP} framework.
 In Section \ref{strategy2}, we describe yet another method
that relies on cuts in $n$-dimensions. This second method is proven
more suitable for numerical applications within the {\tt OPP} algorithm
and may also be applied in a more general framework. Finally, in the
last section, we outline our conclusions.
%------------------------------------------------------------
%------------------------------------------------------------
\section{The origin of ${\rm R_2}$ \label{origin}}
Our starting point is the general expression for the
{\it integrand} of a generic $m$-point
one-loop (sub-)amplitude~\cite{Ossola:2006us}
\bqa
\label{eq:1}
\bar A(\bar q)= \frac{\bar N(\bar q)}{\db{0}\db{1}\cdots \db{m-1}}\,,~~~
\db{i} = ({\bar q} + p_i)^2-m_i^2\,,~~~ p_0 \ne 0\,.
\eqa
In the previous equation, dimensional regularization is assumed, so that
we use a bar to denote objects living
in $n=~4+\epsilon$  dimensions. Furthermore $\bar q^2= q^2+ \tld{q}^2$, where
$\tld{q}^2$ is $\epsilon$-dimensional and $(\tld{q} \cdot q) = 0$.
The numerator function $\bar{N}(\bar q)$ can be also
split into a $4$-dimensional plus a $\epsilon$-dimensional part
\bqa
\label{eq:split}
\bar{N}(\bar q) = N(q) + \tld{N}(\tld{q}^2,q,\epsilon)\,.
\eqa
$N(q)$ is $4$-dimensional
(and will be discussed in the next section) while
$\tld{N}(\tld{q}^2,q,\epsilon)$ gives rise to the RTs of kind ${\rm R_2}$,
defined as
\bqa
\label{eqr2}
{\rm R_2} \equiv  \frac{1}{(2 \pi)^4}\int d^n\,\bar q
\frac{\tld{N}(\tld{q}^2,q,\epsilon)}{\db{0}\db{1}\cdots \db{m-1}}
\equiv  \frac{1}{(2 \pi)^4} \int d^n\,\bar q \,{\cal R}_2\,.
\eqa
To investigate the explicit form of $\tld{N}(\tld{q}^2,q,\epsilon)$
it is important to understand better the separation in
\eqn{eq:split}.
From a given {{\em integrand} $\bar A (\bar q)$
this is obtained by splitting, in the numerator function, the $n$-dimensional integration momentum
${\bar q}$, the $n$-dimensional $\gamma$ matrices
$\bar \gamma_{\bar \mu}$  and the $n$-dimensional metric tensor
$\bar g^{\bar \mu \bar \nu}$ into a $4$-dimensional
component plus remaining pieces:
\bqa
\label{qandg}
\bar q                 &=& q + \tld{q}\,, \nl
\bar \gamma_{\bar \mu} &=&  \gamma_{\mu}+ \tld{\gamma}_{\tld{\mu}}\,,\nl
 \bar g^{\bar \mu \bar \nu}  &=&  g^{\mu \nu}+  \tld{g}^{\tld{\mu} \tld{\nu}}\,.
\eqa
Notice that, when a $n$-dimensional index
is contracted with a 4-dimensional (observable) vector
$v_\mu$, the $4$-dimensional part is automatically selected. For example
\bqa
\bar q \cdot v= q \cdot v\,~~{\rm and}~~\rlap/ {\bar v}= \rlap /v\,.
\eqa

A practical way to compute ${\rm R_2}$ is determining, once for all,
tree-level like Feynman Rules for the theory at hand by
calculating, with the help \eqn{qandg}, the ${\rm R_2}$ part coming from
one-particle irreducible amplitudes up to four external legs.
The fact that four external legs are enough is guaranteed
by the ultraviolet nature of the RTs, proven in~\cite{directcomp1}.

As an illustrative example, we derive
the complete set of the Feynman Rules needed in QED. Along such a line
${\rm R_2}$ can be straightforwardly computed in any theory.
We start from the one-loop $\gamma e^+ e^-$ amplitude
in Fig.~\ref{fig1}.
%----------------------------------------------------------------
\begin{figure}[h]
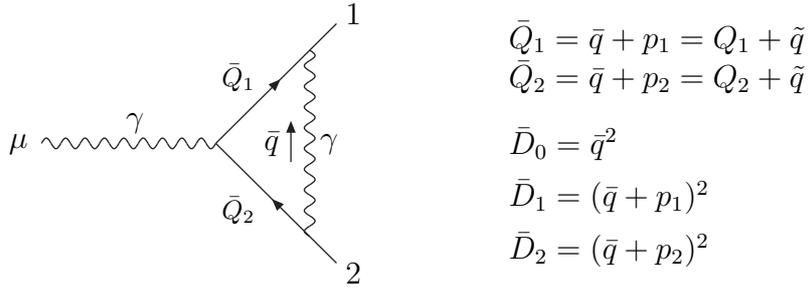

\bfig(300,130)
\SetScale{1}
\sof(-10,00)
\flin{130,5}{85,50}\flin{85,50}{130,95}
\glin{20,50}{85,50}{9}
\glin{120,15}{120,85}{9}
\LongArrow(113,43)(113,57)
\Text(109,50)[r]{$\bar q$}
\Text(15,50)[r]{$\mu$}
\Text(131,50)[r]{$\gamma$}
\Text(55,54)[b]{$\gamma$}
\Text(134,95)[bl]{$1$}
\Text(134,5)[tl]{$2$}
\Text(100,30)[tr]{\footnotesize{$\bar {Q}_2$}}
\Text(100,70)[br]{\footnotesize{$\bar {Q}_1$}}
\Text(195,90)[l]{$\bar Q_1= \bar q+p_1= Q_1 + \tld{q}$}
\Text(195,75)[l]{$\bar Q_2= \bar q+p_2= Q_2 + \tld{q}$}
\Text(195,50)[l]{$\bar D_0= \bar{q}^2$}
\Text(195,30)[l]{$\bar D_1= (\bar q+p_1)^2$}
\Text(195,10)[l]{$\bar D_2= (\bar q+p_2)^2$}
\end{picture}
\end{center}
\caption{QED $\gamma e^+ e^-$ diagram in $n$ dimensions.}
\label{fig1}
\end{figure}
The numerator can be written as follows
\bqa
{\bar N}(\bar q) &\equiv& e^3\, \left\{
     \bar \gamma_{\bar{\beta}}\,\,
      (\rlap/\bar{Q}_1+m_e)
      \,\,\gamma_\mu \,\,
      (\rlap/\bar{Q}_2+m_e)\,\,
     \bar \gamma^{\bar{\beta}} \right\} \nl
&=& e^3\, \left\{
 \gamma_\beta (\rlap/Q_1+m_e) \gamma_\mu (\rlap/Q_2+m_e)
    \gamma^\beta \right. \nl
&-& \left.
\epsilon\,    (\rlap/Q_1-m_e) \gamma_\mu (\rlap/Q_2-m_e)
+ \epsilon \tilde{q}^2\, \gamma_\mu
-\tilde{q}^2\, \gamma_\beta \gamma_\mu \gamma^\beta \right\}\,,
\label{eqver}
\eqa
where all $\epsilon$-dimensional $\gamma$-algebra has been
explicitly worked out
\footnote{$\epsilon$-dimensional $\gamma$ matrices freely anti-commute
with four-dimensional ones: $\{\gamma_\mu,\tilde{\gamma}_\nu\} \,\,= 0$.}
in order to get the desired splitting.
The first term in the l.h.s of \eqn{eqver} is $N(q)$,
while the sum of the remaining
three define $\tld{N}(\tld{q}^2,q,\epsilon)$ for the case at hand.
By inserting $\tld{N}(\tld{q}^2,q,\epsilon)$ in \eqn{eqr2} and
using the fact that
\bqa
\int d^n \bar{q}
\frac{\tld{q}^2}{\db{0}\db{1}\db{2}}       &=& - \frac{i \pi^2}{2} +
\cal{O}(\epsilon)\,,\nl
\int d^n \bar{q}
\frac{q_\mu q_\nu}{\db{0}\db{1}\db{2}}     &=& - \frac{i \pi^2}{2\epsilon}
g_{\mu\nu} + {\cal O}(1)\,,
\eqa
gives
\bqa
{\rm R_2}= -\frac{i e^3}{8 \pi^2} \gamma_\mu + \cal{O}(\epsilon) \,,
\eqa
that can be used to define the following effective vertex:

\begin{figure}[h]
\bfig(300,60)
\SetScale{0.5}
\sof(30,-10)
\flin{130,5}{85,50}\flin{85,50}{130,95}
\glin{20,50}{85,50}{9}
\Text(7,20)[r]{\small $\mu $}
\Text(42.5,25)[]{\Large $\bullet$}
\Text(80,25)[l]{$\displaystyle = -\frac{i e^3}{8 \pi^2} \gamma_\mu$}
\end{picture}
\end{center}
\caption{QED $\gamma e^+ e^-$ effective vertex contributing to ${\rm R_2}$.}
\label{fig2}
\end{figure}

With analogous techniques, taking also into account the integrals given
in \eqn{eq:ratexp}, one gets all the remaining QED effective vertices
given in Fig.~\ref{fig3}.

\begin{figure}[b]
\bfig(300,150)
\SetScale{0.5}
\sof(30,100)
\LongArrow(47,59)(73,59)
\Text(27,32)[bl]{$p$}
\glin{35,50}{85,50}{7}
\Text(42.5,25)[]{\Large $\bullet$}
\glin{85,50}{135,50}{7}
\Text(15,20)[r]{\small $\mu $}
\Text(68,20)[l]{\small $\nu $}
\Text(80,25)[l]{$\displaystyle = -\frac{i e^2}{8 \pi^2}\,
g_{\mu \nu}\, (2 m_e^2-p^2/3)$}
\sof(30,50)
\LongArrow(47,59)(73,59)
\Text(27,32)[bl]{$p$}
\flin{35,50}{85,50}
\Text(42.5,25)[]{\Large $\bullet$}
\flin{85,50}{135,50}
\Text(80,25)[l]{$\displaystyle = \frac{i e^2}{16 \pi^2} (-\rlap/p + 2 m_e)$}
\sof(30,0)
\glin{130,5}{85,50}{7}\glin{85,50}{130,95}{7}
\glin{40,5}{85,50}{7}\glin{85,50}{40,95}{7}
\Text(68,0)[l]{\small $\rho $}
\Text(17,0)[r]{\small $\sigma $}
\Text(68,50)[l]{\small $\nu $}
\Text(17,50)[r]{\small $\mu $}
\Text(42.5,25)[]{\Large $\bullet$}
\Text(80,25)[l]{$\displaystyle = \frac{i e^4}{12 \pi^2}\,
(g_{\mu \nu} g_{\rho \sigma}+g_{\mu \rho} g_{\nu \sigma}+g_{\mu \sigma} g_{\nu \rho})$}
\end{picture}
\end{center}
\caption{QED $\gamma\gamma$, $e e$ and $\gamma \gamma \gamma \gamma$  effective vertices contributing to ${\rm R_2}$.}
\label{fig3}
\end{figure}

To summarize, the problem of computing ${\rm R_2}$ is reduced to a
tree level calculation and we consider it fully solved. The ${\rm
R_1}$ part is, instead, deeply connected to the structure of the
one-loop amplitude, as we shall see in the next section. It is
worthwhile to mention that only the full ${\rm R=R_1+R_2}$
constitutes a physical gauge-invariant quantity in dimensional
regularization.

\section{The {\tt OPP} method and the origin of ${\rm R_1}$
\label{strategy1}}
The {\tt OPP} reduction algorithm provides a useful framework to understand
the origin of the RTs of kind ${\rm R_1}$.
The starting point of the method is an expansion of $N(q)$ in terms
of $4$-dimensional denominators
$\d{i} = ({q} + p_i)^2-m_i^2$
\bqa
\label{eq:2}
N(q) &=&
\sum_{i_0 < i_1 < i_2 < i_3}^{m-1}
\left[
          d( i_0 i_1 i_2 i_3 ) +
     \tld{d}(q;i_0 i_1 i_2 i_3)
\right]
\prod_{i \ne i_0, i_1, i_2, i_3}^{m-1} \d{i} \nl
     &+&
\sum_{i_0 < i_1 < i_2 }^{m-1}
\left[
          c( i_0 i_1 i_2) +
     \tld{c}(q;i_0 i_1 i_2)
\right]
\prod_{i \ne i_0, i_1, i_2}^{m-1} \d{i} \nl
     &+&
\sum_{i_0 < i_1 }^{m-1}
\left[
          b(i_0 i_1) +
     \tld{b}(q;i_0 i_1)
\right]
\prod_{i \ne i_0, i_1}^{m-1} \d{i} \nl
     &+&
\sum_{i_0}^{m-1}
\left[
          a(i_0) +
     \tld{a}(q;i_0)
\right]
\prod_{i \ne i_0}^{m-1} \d{i} \nl
     &+& \tld{P}(q)
\prod_{i}^{m-1} \d{i}\,. \eqa
Inserted back in \eqn{eq:1}, this expression
simply states the multi-pole nature of any $m$-point one-loop amplitude.
The fact that only terms up to 4 poles appear
is due to the fact that $m$-point scalar loop functions with $m > 4$
are always expressible in terms of boxes up
to contributions ${\cal O}(\epsilon)$.
The last term with no poles, $\tld{P}(q)$, has been inserted for generality,
but is zero in practical calculations where $m$-point amplitudes behave
such as $N( \lambda q) \to \lambda^m$ when $\lambda \to \infty$.
The coefficients of the poles can be further split in two pieces.
A piece that still depend on $q$ (the terms
$\tld{d},\tld{c},\tld{b},\tld{a}$), that vanishes upon integration
due to Lorentz invariance, and a piece
that do not depend on q (the terms $d,c,b,a$).
 Such a separation is always possible, as shown in~\cite{Ossola:2006us}, and, with
this choice, the latter set of coefficients is therefore immediately
interpretable as the ensemble of the
coefficients of all possible 4, 3, 2, 1-point
one-loop functions contributing to the amplitude.

 Once \eqn{eq:2} is established, the task of computing the one-loop amplitude
is then reduced, in the {\tt OPP} method, to the algebraical problem of fitting
the coefficients $d,c,b,a$ by evaluating the function $N(q)$
a sufficient number of times, at different values of $q$,
and then inverting the system. Notice that this can be performed
{\em at the amplitude level} and that one does not need to
repeat the work for all Feynman diagrams, provided their sum is known.

 The {\tt OPP} expansion is written in terms of 4-dimensional denominators.
On the other hand, $n$-dimensional denominators $\db{i}$ appear in
\eqn{eq:1}, that differ by an amount $\tld{q}^2$ from their
$4$-dimensional counterparts \bqa \label{relddb} \db{i}= \d{i} +
\tld{q}^2\, \eqa The result of this is a mismatch in the
cancellation of the $n$-dimensional denominators of \eqn{eq:1} with
the $4$-dimensional ones of \eqn{eq:2} (the {\tt OPP} expansion),
that originates a Rational Part. In fact, by inserting \eqn{relddb}
into \eqn{eq:2}, one can rewrite it in terms of $n$-dimensional
denominators (therefore restoring the exact cancellation), but at
the price of adding an extra piece $f(\tld{q}^2,q)$. The RTs of kind
${\rm R_1}$ are defined as \bqa {\rm R_1} \equiv \frac{1}{(2 \pi)^4}
\int d^n\,\bar q \frac{f(\tld{q}^2,q)}{\db{0}\db{1}\cdots
\db{m-1}}\,. \eqa

The explicit form of the function  $f(\tld{q}^2,q)$ can be easily
and explicitly obtained, in the framework of the {\tt OPP} method,
by rewriting any denominator appearing
in \eqn{eq:1} as follows
\bqa
\frac{1}{\db{i}} = \frac{\zb{i}}{\d{i}}\,,~~~~{\rm with}~~~
\zb{i}\equiv \left(1- \frac{\tld{q}^2}{\db{i}} \right)\,.
\eqa
This results in
\bqa
\label{eq:3}
\bar A(\bar q)= \frac{N(q)}{\d{0}\d{1}\cdots \d{m-1}}\,
\zb{0} \zb{1} \cdots \zb{m-1} + {\cal R}_2\, ,
\eqa
where ${\cal R}_2$ is the integrand function introduced in \eqn{eqr2}.
Then, by inserting \eqn{eq:2} in \eqn{eq:3}, one obtains
\bqa
\label{eq:4}
\bar A(\bar q) &=&
\sum_{i_0 < i_1 < i_2 < i_3}^{m-1}
\frac{
          d( i_0 i_1 i_2 i_3 ) +
     \tld{d}(q;i_0 i_1 i_2 i_3)
}{\db{i_0} \db{i_1} \db{i_2} \db{i_3}}
\prod_{i \ne i_0, i_1, i_2, i_3}^{m-1} \zb{i} \nl
     &+&
\sum_{i_0 < i_1 < i_2 }^{m-1}
\frac{
          c( i_0 i_1 i_2) +
     \tld{c}(q;i_0 i_1 i_2)
}{\db{i_0} \db{i_1} \db{i_2}}
\prod_{i \ne i_0, i_1, i_2}^{m-1} \zb{i} \nl
     &+&
\sum_{i_0 < i_1 }^{m-1}
\frac{
          b(i_0 i_1) +
     \tld{b}(q;i_0 i_1)
}{\db{i_0} \db{i_1}}
\prod_{i \ne i_0, i_1}^{m-1} \zb{i} \nl
     &+&
\sum_{i_0}^{m-1}
\frac{
          a(i_0) +
     \tld{a}(q;i_0)
}{\db{i_0}}
\prod_{i \ne i_0}^{m-1} \zb{i} \nl
     &+& \tld{P}(q)
\prod_{i}^{m-1} \zb{i}+{\cal R}_2\,.
\eqa
${\rm R_1}$ is then produced,
after integrating over $d^n \bar q$, by the
$\tld{q}^2$ dependence coming from the various $\zb{i}$ in \eqn{eq:4}.
This strategy have been adopted in~\cite{sixphotons}, where also all
needed integrals have been carefully classified and computed.

Although rather transparent, the above derivation of ${\rm R_1}$ has
two drawbacks. First of all, it requires the knowledge of the
spurious terms \footnote{After multiplication with the $\zb{i}$,
they give non vanishing contributions.}. Secondly, it is not
suitable when combining diagrams together because, when taking
common denominators, additional terms containing $\tld{q}^2$ appear
in the numerator, that may give rise to new rational parts. The
bookkeeping of such new structures is equivalent to the treatment of
each diagram separately, jeopardizing the ability of the
4-dimensional {\tt OPP} technique of dealing directly with the
amplitude. For these reasons we present, in the next section, a
different way of attacking this problem that does not relies on
spurious terms and that also allows one to combine diagrams {\em
before} fitting the coefficients $d,c,b,a$. This second method is
better suited for a numerical implementation, and it has been
already successfully implemented in a Fortran
code~\cite{Ossola:2007ax}.

\section{The $n$-dimensional cuttings to compute ${\rm R_1}$ \label{strategy2}}
The Rational Terms ${\rm R_1}$ can be computed by looking at the implicit mass
dependence (namely reconstructing powers of $\tld{q}^2$) in
the coefficients $d,c,b$ of the one-loop functions,
once $\tld{q}^2$ is reintroduced through the
mass shift
\bqa
\label{massshift}
m_i^2 \to m_i^2 -\tld{q}^2\,.
\eqa
This procedure is formally equivalent, in the generalized unitarity
framework, to the applications of $n$-dimensional cuts, and is obtained,
in the {\tt OPP} language, by simply performing the {\tt OPP}
expansion of \eqn{eq:2} directly in
terms of the $n$-dimensional denominators of \eqn{relddb}.
 By doing so, all coefficients of the {\tt OPP} expansion start depending on
$\tld{q}^2$. The spurious terms keep being spurious, because they
vanish due to Lorentz invariance (that is untouched when
including powers of $\tld{q}^2$), while the coefficients
$d,c,b$ generate the following extra integrals~\cite{Ossola:2006us}
\bqa \label{eq:ratexp}
\int d^n \bar{q}
\frac{\tld{q}^2}{\db{i}\db{j}}             &=& - \frac{i \pi^2}{2}
\left[m_i^2+m_j^2-\frac{(p_i-p_j)^2}{3} \right]   +
\cal{O}(\epsilon)\,, \nl
\int d^n \bar{q}
\frac{\tld{q}^2}{\db{i}\db{j}\db{k}}       &=& - \frac{i \pi^2}{2} +
\cal{O}(\epsilon)\,,\nl
\int d^n \bar{q}
\frac{\tld{q}^4}{\db{i}\db{j}\db{k} \db{l}} &=& - \frac{i \pi^2}{6} +
\cal{O}(\epsilon)\,.\nl
\eqa
One can prove that
\bqa
\label{bandc}
b(ij;\tld{q}^2) &=&   b(ij)
                     +\tld{q}^2 b^{(2)}(ij) \,, \nl
c(ijk;\tld{q}^2) &=&   c(ijk)
                     +\tld{q}^2 c^{(2)}(ijk)\,.
\eqa
Furthermore, by using \eqn{massshift}, the first line of \eqn{eq:2}
becomes
\bqa
\label{bigd0}
{\cal D}^{(m)}(q,\tld{q}^2) \equiv \sum_{i_0 < i_1 < i_2 < i_3}^{m-1}
\left[
          d( i_0 i_1 i_2 i_3 ;\tld{q}^2) +
     \tld{d}(q;i_0 i_1 i_2 i_3;\tld{q}^2)
\right]
\prod_{i \ne i_0, i_1, i_2, i_3}^{m-1} \db{i} \,,
\eqa
and the following expansion holds
\bqa
\label{bigd}
{\cal D}^{(m)}(q,\tld{q}^2)= \sum_{j= 2}^{m} \tld{q}^{(2j-4)} d^{(2j-4)}(q)\,,
\eqa
where the last coefficient is independent on $q$
\bqa
\label{noq}
d^{(2m-4)}(q) = d^{(2m-4)}\,.
\eqa
In practice, once the $4$-dimensional
coefficients have been determined, one simply redoes
the fits for different values of $\tld{q}^2$, in order to determine
$b^{(2)}(ij)$, $c^{(2)}(ijk)$ and  $d^{(2m-4)}$.
Such three quantities are the coefficients of the three
extra scalar integrals listed in \eqn{eq:ratexp}, respectively, so
that
\bqa
\label{erre1}
{\rm R}_1 &=& - \frac{i}{96 \pi^2} d^{(2m-4)}
      -  \frac{i}{32 \pi^2}
\sum_{i_0 < i_1 < i_2 }^{m-1}
          c^{(2)}( i_0 i_1 i_2)
 \nl
     &-& \frac{i}{32 \pi^2}
\sum_{i_0 < i_1 }^{m-1}
          b^{(2)}(i_0 i_1)
\left(m_{i_0}^2+m_{i_1}^2- \frac{(p_{i_0}-p_{i_1})^2}{3}\right )\,. \eqa
In \app{appa}, we prove \eqnss{bandc}{noq}, we single out the origin
of $d^{(2m-4)}$ as the coefficient of the last
integral of \eqn{eq:ratexp} and we show how it can be also
derived outside the {\tt OPP} technique.
Finally, yet another way of computing $ d^{(2m-4)}$ can be obtained
by noticing that
\bqa
\label{eqd}
d^{(2m-4)} &=&  \lim_{\tld{q}^2 \to \infty}
\frac{{\cal D}^{(m)}(q,\tld{q}^2)}{\tld{q}^{(2m-4)}}\,.
\eqa
This last method is also implemented in the code of Ref.~\cite{Ossola:2007ax}.

We close this section by stressing that the way of computing
the coefficients appearing in \eqn{erre1} is immaterial. Therefore
the method to extract ${\rm R_1}$ described in this section, namely by looking at
the mass dependence of the coefficients of the scalar loop functions,
can be used independently on the {\tt OPP} technique.
In particular, one can derive all needed coefficients
also with the help of analytical methods.
\section{Conclusions\label{conclusions}}
We have discussed and clarified the origin of the Rational Terms appearing in
one-loop amplitudes, showing that they can be classified in two classes.
The first class ($\rm R_2$) can be computed by defining
tree-level like Feynman rules for the theory at hand.
We precisely outlined the way to derive the needed extra Feynman rules,
listing them explicitly in the case of QED. We therefore consider the
problem of computing $\rm R_2$ completely solved.
The second piece ($\rm R_1$) can be calculated in two different ways.
We presented a first technique that relies on the
{\tt OPP} method and a second, more general, computational strategy.
Both methods have been successfully tested within the {\tt OPP}
method. The second one, however, is more suitable for a numerical
implementation, and it has been used in the numerical code {\tt
CutTools}.
%This second way is more suitable for a numerical implementation,
%and it has been already successfully used in a numerical code.

\section*{Acknowledgments}

G.O. and R.P. acknowledge
the financial support of the ToK Program ``ALGOTOOLS'' (MTKD-CD-2004-014319).

C.G.P.'s and R.P.'s research was partially supported by the RTN
European Programme MRTN-CT-2006-035505 (HEPTOOLS, Tools and Precision
Calculations for Physics Discoveries at Colliders).

The research of R.P. was also supported by MIUR under contract
2006020509\_004 and by the MEC project FPA2006-05294.

C.G.P. and R.P. thank the  Galileo Galilei Institute for Theoretical
Physics for the hospitality and the INFN for partial support during the
completion of this work.

\section*{Appendices}
\appendix
\section{The $\tld{q}^2$ dependence of the {\tt OPP} coefficients \label{appa}}
Our starting point is a rank $r$ tensor $m$-point {\em integrand}
defined as
\bqa
A_{m;r} \equiv \frac{q_{\mu_1} \cdots q_{\mu_r}}{\db{0} \cdots \db{m-1}}
\,~~~(m > 3\,,~r > 1)\,.
\eqa

By expressing the integration momentum $q$ in terms of the basis of
the external vectors, with coefficients linearly depending on the
propagator function appearing in the
denominator\cite{Ossola:2006us}, one ends up in an expression that
it is identical to the {\tt OPP} master equation plus terms
containing higher-point scalar amplitudes $A_{k;0}$ $k=5,\ldots,m$.
Since all reductions formula used so far are simple polynomial in
terms of $\tilde{q}^2$, the same is true for all coefficients
appearing in that expansion. This, in conjunction with the fact that
the maximum rank allowed $r=m$, easily proves first of all
\eqn{bandc}. In the next step one reduces the $k-$point scalar terms
$k=5,\ldots,m$ in terms of the $4-$point ones at the integrand
level. After that step the individual $d$ and $\tilde{d}$
coefficients become rational functions of $\tilde{q}^2$.
Nevertheless, since the ${\cal D}^{(m)}(q,\tld{q}^2)$ of
\eqns{bigd0}{bigd} is nothing more than the combined numerator of
all scalar terms with $k=4,\ldots,m$, and all coefficients are
polynomials in $\tilde{q}^2$, so is the ${\cal
D}^{(m)}(q,\tld{q}^2)$ function. Finally it is straightforward to
see that in the {\tt OPP} expansion in terms of $n-$dimensional
propagators, the $\tilde{q}^2\to \infty$ behavior of the individual
$d$ terms is $\tilde{q}^4$, which means that the only integral
involved is the last one of \eqn{eq:ratexp}.

Another way to derive the same results is
by using the reduction at the {\em integrand} level introduced in
~\cite{intlevel}.
One can express $A_{m;r}$ as a linear combination, with tensor coefficients,
of five classes of lower rank tensors:
$A_{m;r-1},\tld{q}^2A_{m;r-2}, A_{m;r-2}, A_{m-1;r-1}$ and $A_{m-1;r-2}$.
To keep things as transparent as possible, we omit explicitly writing
the coefficients and we denote such a linear combination using 
the following notation
\footnote{Notice that each of the five terms of \eqn{comb1} may
actually represent an entire class of contributions with
different combinations of denominators. For example
$A_{m-1;r-1}$ stands for all $m$ rank $r-1$ tensor {\em integrands}  
that can be obtained by omitting $1$ among the original $m$ possible denominators.}
\bqa
\label{comb1}
A_{m;r}=
\left\{
A_{m;r-1}           |
\tld{q}^2 A_{m;r-2} |
          A_{m;r-2} |
A_{m-1;r-1} |
A_{m-1;r-2}
\right\}\,.
\eqa
Analogously, it was proven that
\bqa
\label{comb2}
A_{m;1} &=&
\left\{
A_{m;0}  |
A_{m-1;0}
\right\}\, \,~~~(m > 4)\,, \nl
A_{4;1} &=&
\left\{
A_{4;0}  |
A_{3;0}  |
\tld{d}(q) A_{4;0}
\right\}\,,
\eqa
where $\tld{d}(q)$ is defined such as
\bqa
\int d^n\,\bar q\, \tld{d}(q) A_{4;0}  =0\,.
\eqa

By using \eqns{comb1}{comb2} it is easy to constructively prove
\eqns{bigd}{noq}. We explicitly give the derivation for the case
$m=6$ and $r=6$.
By iteratively applying \eqns{comb1}{comb2}, one ends up with
\bqa
\label{expr}
A_{6;6} &=&
\sum_{j= 4,5,6}
\left\{
A_{j;0}            |
\tld{q}^2 A_{j;0}  |
\tld{q}^4 A_{j;0}
\right\}
+
\left\{
\tld{q}^6 A_{6;0}
\right\}
+
\left\{
\tld{d}(q) A_{4;0}  |
\tld{d}(q)\tld{q}^2 A_{4;0}
\right\} \nl
&+&
{\cal O}(A_{3;r_3})\,,
\eqa
where ${\cal O}(A_{3;r_3})$ means that we are neglecting contributions
with 3 or less denominators.
By power counting, only the term $\tld{q}^4 A_{4;0}$ contributes to $R_1$.
Notice also that its coefficient (that we call $z_4$) is independent on $q$.
Now we can take a common denominator in \eqn{expr} by multiplying and
dividing 5 and 4-point structures by the relevant missing $n$-dimensional
propagators.
In particular, for example, by calling $\db{i}$ and  $\db{j}$ the 2 
denominators that do not appear in $A_{4;0}$
\bqa
\label{eqz4}
z_4\, \tld{q}^4 A_{4;0}= 
 z_4 \tld{q}^4 \db{i} \db{j} A_{6;0} =
 z_4 \tld{q}^4 (\tld{q}^2+D_i)(\tld{q}^2+D_j) A_{6;0}\,.
\eqa
The numerator of the resulting expression is polynomial in $\tld{q}^2$ and
it is nothing but the function
${\cal D}^{(m)}(q,\tld{q}^2)$ of \eqns{bigd0}{bigd}, with $m= 6$. Furthermore
$d^{(8)}= z_4$, independent on q.
The general case can be derived along the same lines.

\eqn{eqz4} also clarifies why $d^{(2m-4)}$ is the coefficient of the 
4-point like last integrals of \eqn{eq:ratexp}}. In fact,
$m-4$ among the $m$ original $n$-dimensional denominators
always completely factorize in front of $d^{(2m-4)}$. 
Notice also that the origin of the coefficient $d^{(2m-4)}$ is uniquely coming
from the mass dependence of the coefficients of the 4-point scalar functions
after tensor reduction, but {\em before expressing $m$-point scalar functions with $m>4$ in terms of boxes}. The reason why we do not
reduce \eqn{expr} to structures
with 4-denominators is that
this would bring a $\tld{q}^2$ dependence in the denominator,
when passing from 5 to 4 denominators. In our notation~\cite{fivepoint}
\bqa
A_{5;0} =
\left\{
\frac{1}{\tld{q}^2+c_i} A_{4;0}  \right| \left.
\frac{1}{\tld{q}^2+c_i} \tld{d}(q) A_{4;0}
\right\}\,,
\eqa
with $c_i$ constants.
It is therefore much better to take, instead, common denominators.
Finally, analogous techniques can be used to prove \eqn{bandc}.

\end{document}